\documentclass[%
reprint,
amsmath,amssymb,
aps, superscriptaddress
]{revtex4-1}

\usepackage{graphicx}
\usepackage{dcolumn}
\usepackage{bm}
\usepackage{array}
\usepackage{hyperref}
\hypersetup{colorlinks,%
	citecolor=blue,%
	urlcolor=blue}
\usepackage{subfigure}

\begin{document}


\title{Spin-orbit-stable type-II nodal line band crossing in $n$-doped monolayer MoX${}_{2}$(X=S,Se,Te)}%

\author{Kyung-Han Kim}
\affiliation{Department of Physics, Pohang University of Science and Technology, Pohang 37673, Korea}


\author{Bohm Jung Yang}
\affiliation{Department of Physics and Astronomy, Seoul National University, Seoul 08826, Korea}
\affiliation{Center for Correlated Electron Systems, Institute for Basic Science (IBS), Seoul 08826, Korea}
\affiliation{Center for Theoretical Physics (CTP), Seoul National University, Seoul 08826, Korea}

\author{Hyun-Woo Lee}
\email{hwl@postech.ac.kr}
\affiliation{Department of Physics, Pohang University of Science and Technology, Pohang 37673, Korea}




\date{\today}

\begin{abstract}

1H-phase monolayer transition-metal dichalcogenides have spin-split electronic band structures near the $K(K')$ point due to strong spin-orbit coupling and intrinsic inversion symmetry breaking. In case of the monolayer MoX${}_{2}$ (X=S, Se, Te), the spin-split conduction bands cross near the $K(K')$ point. We show that the band crossing occurs not only along the high symmetry directions, but also along arbitrary directions due to a mirror reflection symmetry of the monolayer. As a result, $n$-doped monolayer MoX${}_{2}$ is a two-dimensional type-II nodal-line material.

\end{abstract}

\maketitle


\section{Introduction}



Dirac and Weyl semi-metals have been extensively investigated due to their unique topological properties~\cite{N. P. Armitage 2018}.
Recently, nodal line semi-metal has also attracted huge attention as a new kind of topological semi-metal~\cite{S. Yang 2018}.
So far, several materials are proposed as a three-dimensional (3D) nodal line semi-metal such as 3D graphene, Cu${}_{3}$PdN, AX${}_{2}$(A = Ca, Sr, Ba; X = Si, Ge, Sn), CaTe, Mg${}_{3}$Bi${}_{2}$, $\beta$-PbO${}_{2}$, and CaP${}_{3}$ family of materials~\cite{H. Weng 2015,R. Yu 2015,H. Huang 2016,Y. Du 2017,X. Zhang 2017,Z. Wang 2017,Q. Xu 2017}.
These proposals are however based on the assumption that there is no spin-orbit coupling (SOC). When SOC is taken into account, the nodal line structure in the materials is broken, though weakly since SOC is weak.
A recent first-principles calculation~\cite{S. Nie 2018} combined with an effective model analysis predicts that layered ferromagnetic rare-earth-metal monohalides are 3D nodal line semi-metal with SOC.
Until now, PbTaSe$_{2}$ is the only material that is confirmed experimentally as a 3D nodal line semi-metal with strong SOC~\cite{G. Bian 2016}.


The concepts of Dirac, Weyl, and nodal lines are applicable to two-dimension (2D) as well.
Several theoretical calculations~\cite{J. Lu 2015,Y. Jin 2017,P. Zhou 2018} predicted nodal line band crossing in 2D materials with weak SOC.
But similarly to the 3D case, the nodal line band crossing in the materials is broken when SOC is taken into account.
An experiment~\cite{B. Feng 2017} reported the realization of 2D Dirac nodal line fermions in monolayer Cu${}_{2}$Si with weak SOC. When SOC is taken into account theoretically, however, the nodal line band crossing in this material is also broken.
A recent theory~\cite{S. Park 2017} classifies all possible band crossing structures in time-reversal invariant 2D noncentrosymmetric systems {\it with SOC}.
According to the theory, the nodal line crossing can be stable with respect to SOC if materials have proper geometrical symmetries.
Weak SOC materials (Be${}_{2}$C, BeH${}_{2}$)~\cite{B. Yang 2016} and Na${}_{2}$CrBi trilayer~\cite{C. Niu 2017} are theoretically proposed as type-I 2D nodal line materials with SOC.

\begin{figure}[ht] 
	\centering 
	\includegraphics[angle=0, width=8cm]{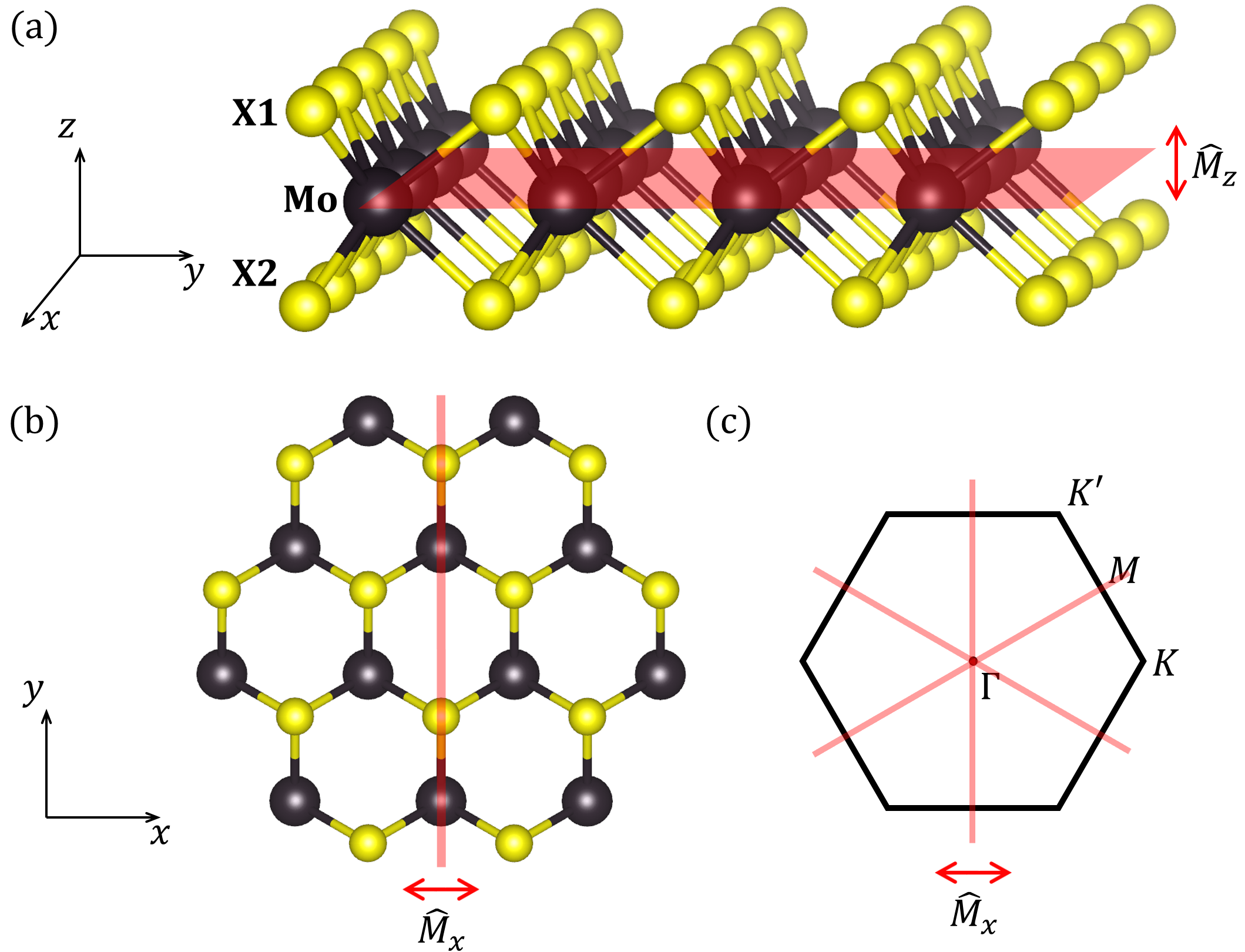} 
	\subfigure{\label{fig:1a}}
	\subfigure{\label{fig:1b}}
	\subfigure{\label{fig:1c}}
	\caption{\label{fig:1}
		(Color online) (a) Side view and (b) Top view of a 1H phase monolayer MoX${}_{2}$~\cite{VESTA}, which consists of Mo atoms (larger dot) and X atoms (smaller dot). It has mirror reflection symmetry with respect to the middle sublayer plane ($\hat{M}_{z}$) and the perpendicular bisector plane ($\hat{M}_{x}$). (c) First Brillouin zone in the momentum space with $\hat{M}_{x}$ mirror symmetry invariant line (red).} 
\end{figure}

In this paper, we report that $n$-doped monolayer MoX${}_{2}$ (X=S, Se, Te) (Fig.~\ref{fig:1}), which is a well-known monolayer transition-metal dichalcogenides (TMDs) material, has a type-II nodal line band crossing that remains robust even though the monolayer has strong SOC. To the best of our knowledge, this is the first example of the 2D type-II nodal line material with strong SOC. The nodal line band crossing is demonstrated by combining tight binding model calculation with symmetry analysis.
It has been already known~\cite{G. B. Liu 2013,H. Rostami 2015,K. Kosmider 2013, A. Kormanyos 2014, A. Kormanyos 2015, J. P. Echeverry 2016, M. Koperski 2017} that the lowest conduction band of the monolayer MoX$_{2}$ has band crossing near the $K(K')$ point along the $\Gamma-K(K')$ line.
The main point of this paper is that the band crossing appears not only along the high symmetry directions, but also along arbitrary directions, thereby forming a nodal line band crossing near the $K(K')$ point.

\begin{figure}[ht] 
	\centering 
	\includegraphics[angle=0, width=8cm]{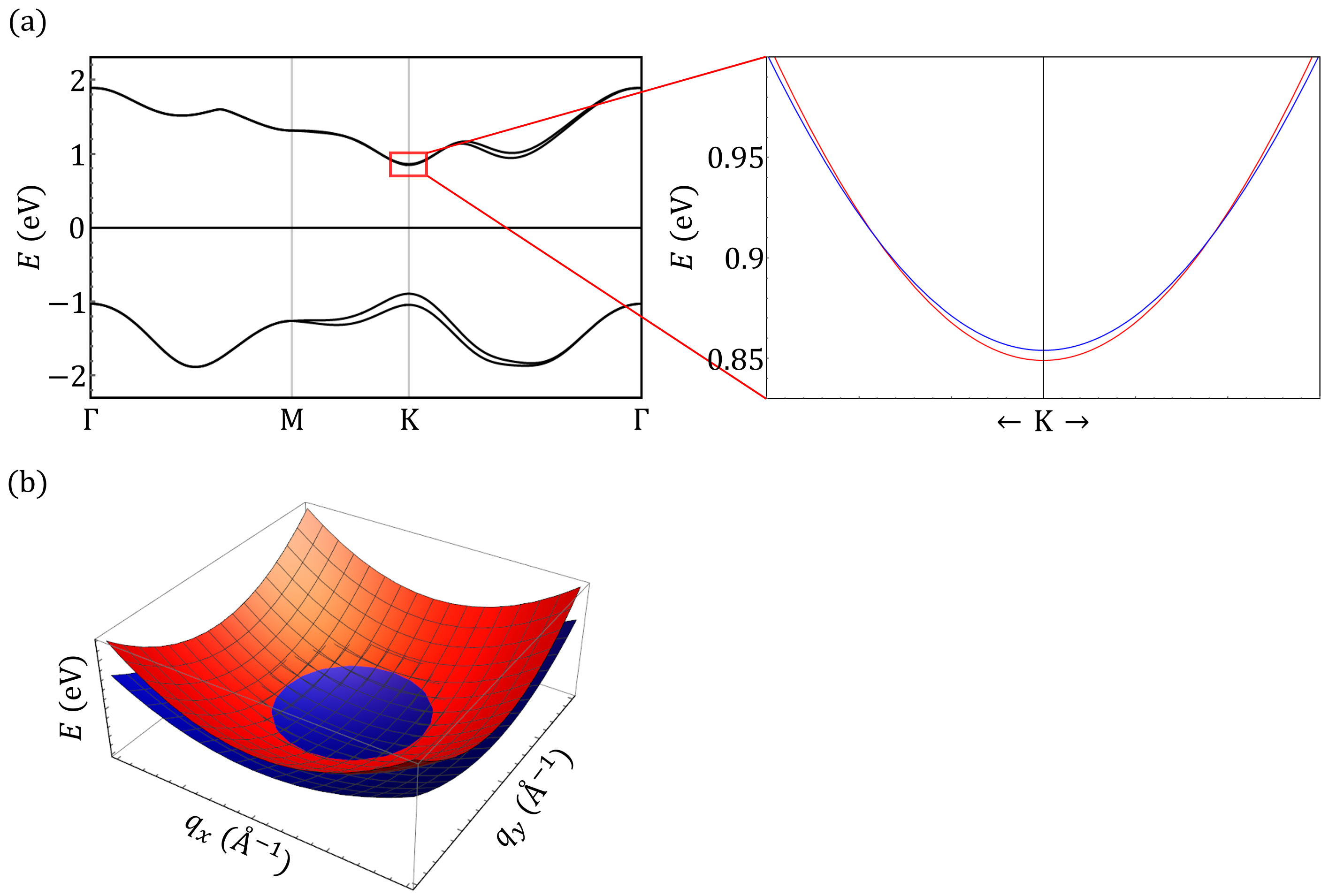} 
	\subfigure{\label{fig:2a}}
	\subfigure{\label{fig:2b}}
	\caption{\label{fig:2}
		(Color online) (a) Tight binding band structure of conduction and valence bands of the 1H phase monolayer MoS${}_{2}$ with the SOC. In the rectangular region near the K point, there is band crossing of the conduction bands which makes a nodal line band structure. Here the colors indicate states with the out-of-plane spin up (red) and down (blue), respectively.
		(b) Schematic 3D plot of the conduction bands near the $K$ point ($\bm{q}=\bm{k}-\bm{K}=0$). It shows the nodal line band crossing between the spin up and down bands. Trigonal warping effect is too weak to be seen near the conduction band minimum.} 
\end{figure}

\section{Theory}

In this Section, we demonstrate that the 1H phase monolayer MoX$_{2}$ has the nodal line band crossing and that the crossing remains stable despite SOC due to its $\hat{M}_{z}$ mirror reflection symmetry~[Fig.~\ref{fig:1a}].
We use the tight binding Hamiltonian of the 1H phase monolayer MoS${}_{2}$~\cite{H. Rostami 2015} to calculate the energy band structure, which is shown along the $\Gamma-K$ line in Fig.~\ref{fig:2a}.
As illustrated in Fig.~\ref{fig:2}, the spin up conduction band and the spin down conduction band cross near the $K$ point and forms a type-II nodal line band crossing.

To check if the structure in Fig.~\ref{fig:2b} is a true nodal line crossing structure or an anti-crossing structure with small energy separation, we utilize a recent theoretical result~\cite{S. Park 2017}, which classifies band crossings and emergent semimetals in two-dimensional noncentrosymmetric systems.
According to the classification, nodal line semimetal phase is topologically stable in 2D material if (i) there exists mirror reflection symmetry about the material's 2D plane and (ii) the two bands which are crossing each other have different mirror eigenvalues.

In case of the 1H phase monolayer MoX${}_{2}$, it is invariant under the mirror reflection about the $xy$-plane (Fig.~\ref{fig:1}), so the condition (i) is satisfied. Then the Hamiltonian $H$ of the monolayer should commute with the mirror reflection operator $\hat{M}_z$,
\begin{equation}
\big[H,\hat{M}_{z}\big] = 0.
\end{equation}
Since $\hat{M}_z$ does not modify the 2D crystal momentum $\bf{k}$, this commutation relation implies that
energy eigenstates with the crystal momentum $\bf{k}$ are eigenstates of $\hat{M}_{z}$. Since the eigenvalue spectrum of $\hat{M}_z$ is discrete, the continuous variation of energy eigenstates as a function of $\bf{k}$ within an energy band cannot alter the eigenvalue of $\hat{M}_{z}$ (or mirror eigenvalue). Thus one finds that
each energy band of $H$ has a well defined mirror eigenvalue.

Now, we consider the mirror eigenvalues of the two conduction bands depicted in Fig.~\ref{fig:2}. To check if the two conduction bands satisfy the conduction (ii), crude information on the wavefunction character of the two conduction bands is sufficient since possible mirror eigenvalues are discrete and the mirror eigenvalue of each band can be determined from crude wavefunction characters.
In this regard, the wavefunction character near the $K$ point is sufficient.
The $K$ point has the ${\rm{C}}_{\rm{3h}}$ point group symmetry. Using its irreducible representations~\cite{A. Kormanyos 2013}, the wavefunctions of the two conduction bands at the $K$ point can be expressed as
\begin{equation}
\bigg[\alpha^{\rm{Mo}}\left|\Psi^{\rm{Mo}}_{2,0}\right>+ \frac{\alpha^{\rm{X}}}{\sqrt{2}}\Big(\left|\Psi^{\rm{X1}}_{1,1}\right>+\left|\Psi^{\rm{X2}}_{1,1}\right>\Big)\bigg]
\otimes\left|\chi\right>,
\label{eq:wavefunction-character}
\end{equation}
where $\chi$ is spin up or down states, $\alpha$ is the ratio of each orbital wavefuntion and the superscript indicates at which atom the corresponding wavefunction is localized. X1 and X2 in the superscript denotes upper and lower X atoms. The subscript denotes, on the other hand, spherical harmonics information of the wavefunction.
From the character table of the group ${\rm{C}}_{\rm{3h}}$~\cite{A. Kormanyos 2013}, or just considering the spherical harmonics information in the orbital part wavefunction character of the conduction bands in Eq.~(\ref{eq:wavefunction-character}), one finds that the orbital part wavefunctions of the two conduction bands are even about $\hat{M}_{z}$.
Since wavefunctions consist of orbital and spin parts, one needs to figure out the effect of $\hat{M}_z$ on the spin part wavefunction as well. For this, we utilize the relation $\hat{M}_z=\hat{I}\exp[-i(L_z+S_z)\pi/\hbar]$, where $\hat{I}$ denotes the inversion operator, and $L_z$ ($S_z$) denotes the orbital (spin) angular momentum operator along the out-of-plane direction. Thus one obtains $\hat{M}_z|\uparrow\rangle=-i |\uparrow\rangle$ and $\hat{M}_z|\downarrow\rangle=+i|\downarrow\rangle$. Then one finds
\begin{eqnarray}
\hat{M}_{z}\left|\psi_{\rm{cb}}(\bm{r})\right>\otimes\left|\uparrow\right>&=&-i\left|\psi_{\rm{cb}}(\bm{r})\right>\otimes\left|\uparrow\right>, \\ \hat{M}_{z}\left|\psi_{\rm{cb}}(\bm{r})\right>\otimes\left|\downarrow\right>&=&i\left|\psi_{\rm{cb}}(\bm{r})\right>\otimes\left|\downarrow\right>.
\end{eqnarray}
Since the two conduction bands share the same orbital character but have opposite spin character, this means that the spin up and spin down conduction bands have different mirror eigenvalues ($-i$ and $+i$, respectively). Therefore the two conduction bands satisfy the condition (ii) as well near the $K$ point.
This verifies that there exist band crossing along arbitrary directions near the $K$ point and the resulting nodal line band crossing near the $K$ point is stable topologically.
An exactly same kind of analysis applies to the type-II nodal line structure band crossing near the $K'$ point as well since the wavefunction characters of the conduction band energy eigenstates at the $K'$ point is obtained by complex conjugation of the corresponding wavefunction characters at the $K$ point.

\section{Discussion}

In the previous section, we showed that if the spin split bands are ordered as Fig.~\ref{fig:2} to form band crossing near the $K(K')$ point,
there should appear a true nodal line structure that is stable against anti-crossing since the anti-crossing is forbidden by the $\hat{M}_{z}$ mirror symmetry.
According to first principles calculations~\cite{K. Kosmider 2013, A. Kormanyos 2014, A. Kormanyos 2015, J. P. Echeverry 2016, M. Koperski 2017}, the band splitting between the two conduction bands near the K and K' points are negative (meaning band crossing) for MoX${}_{2}$ (-3 meV for MoS${}_{2}$, -22 meV for MoSe${}_{2}$, -36 meV for MoTe${}_{2}$) and positive (meaning no band crossing) for WX${}_{2}$ (32 meV for WS${}_{2}$, 37 meV for WSe${}_{2}$, 52 meV for WTe${}_{2}$). In case of MoSe${}_{2}$ (WS${}_{2}$, WSe${}_{2}$), optical experiments~\cite{M. Koperski 2017, Z. Ye 2014, A. Arora 2015, G. Wang 2015, M. R. Molas 2017, X. X. Zhang 2017} also support the band crossing (no crossing). Based on these results, we then conclude that there should appear type-II nodal line structure in MoX${}_{2}$. One cautionary remark is in order. In case of MoS${}_{2}$, the band splitting is very small and thus further studies are necessary to confirm the predicted band structure.  
The reason why the conduction band spin split ordering is different between MoX${}_{2}$ and WX${}_{2}$ is that the level spacing between the two conduction bands with opposite spin is affected in different ways in Mo-based and W-based TMD materials by the competition between the first order SOC contribution from the chalcogen atom's orbitals and the second order SOC contribution from the transition-metal's orbitals~\cite{G. B. Liu 2013, A. Kormanyos 2015}.

According to DFT calculation of 1H phase monolayer MoS${}_{2}$~\cite{A. Kormanyos 2015}, the nodal line band crossing occurs about 50 meV above the conduction band minimum and the $\bf{k}$-space area enclosed by the nodal line occupies about 1.5 percent of the first Brillouin zone. In this respect, it is worth noting that even without any intentional doping, monolayer MoS${}_{2}$ is naturally doped~\cite{K. Dolui 2013} due to the interaction between the MoS$_{2}$ monolayer and its substrate. Many experiments find this natural doping to be $n$-type for typical dielectric substrates~\cite{D. Sercombe 2013}. Thus it is plausible that mild additional $n$-doping of the monolayer MoS$_2$ can bring the Fermi energy close to the nodal line crossing.
When such Fermi energy tuning is achieved,
this kind of nodal line band crossing has the potential to show interesting physics.
For instance, it is predicted~\cite{K. Kim 2018,B. T. Zhou 2017,A. S. Patri 2018} that large Berry curvature is generated when the nodal line band crossing is gapped out by the symmetry breaking. By the way, we remark that the type-II nodal line band crossing does not make $n$-dopped monolayer MoS${}_{2}$ a semi-metal, since the crossing does not suppress the density of state in 2D materials.

Now we consider the stability of the nodal line band crossing against the symmetry breaking.
When the $\hat{C}_{3}$ rotational symmetry is broken, the nodal line is still stable due to the $\hat{M}_{z}$ mirror symmetry.
However, when the $\hat{M}_{z}$ mirror symmetry is broken, the nodal line is no longer stable, and may evolve to either type-II Dirac points or anti-crossing structure.
The type-II nodal line structure evolves to type-II Dirac points if some points on the nodal line remain degenerate due to remaining symmetries after the $\hat{M}_{z}$ mirror symmetry breaking. In this case, MoX${}_{2}$ has the $\hat{M}_{x}$ mirror symmetry and this symmetry can guarantee such degeneracy on the $\hat{M}_{x}$ mirror invariant line.
However as shown in~\ref{fig:1b}, \ref{fig:1c}, $K(K')$ point is not on the $\hat{M}_x$ mirror symmetry line which is along the $\Gamma M$ line in the momentum space, so the nodal line band crossing becomes anti-crossing band structure.

In summary, we showed that the 1H phase monolayer MoX${}_{2}$ has nodal line band crossing near the $K(K')$ point and this band crossing is protected along the arbitrary directions by mirror reflection symmetry about a 2D plane.
As a result, $n$-doped monoyaler MoX${}_{2}$ is predicted to be a type-II nodal line material with strong SOC.

We acknowledge fruitful discussion with Youngkuk Kim. K.H.K and H.W.L were supported by the National Research Foundation of Korea (No.2018R1A5A6075964).
B.-J.Y. was supported by the Institute for Basic Science in Korea (Grant No. IBS-R009-D1), Basic Science Research Program through the National Research Foundation of Korea (NRF) (Grant No. 0426-20170012, No.0426-20180011), the POSCO Science Fellowship of POSCO TJ Park Foundation (No.0426-20180002), U.S. Army Research Office under Grant Number W911NF-18-1-0137.


\begin{thebibliography}{99}

\bibitem{N. P. Armitage 2018}
N. P. Armitage, E. J. Mele, and A. Vishwanath,
\href{https://journals.aps.org/rmp/abstract/10.1103/RevModPhys.90.015001}{Rev. Mod.Phys. $\bm{90}$, 015001 (2018).}

\bibitem{S. Yang 2018}
S. Yang, H. Yang, E. Derunova, S. S. P. Parkin, B. Yan, and M. N. Ali,
\href{https://www.tandfonline.com/doi/full/10.1080/23746149.2017.1414631?scroll=top\&needAccess=true}{Advances in Physics: X $\bm{3}$, 1414631 (2018).}


\bibitem{H. Weng 2015}
H. Weng, Y. Liang, Q. Xu, R. Yu, Z. Fang, X. Dai, and Y. Kawazoe,
\href{https://journals.aps.org/prb/abstract/10.1103/PhysRevB.92.045108}{Phys. Rev. B $\bm{92}$, 045108 (2015).}
\bibitem{R. Yu 2015}
R. Yu, H. Weng, Z. Fang, X. Dai, and X. Hu,
\href{https://journals.aps.org/prl/abstract/10.1103/PhysRevLett.115.036807}{Phys. Rev. Lett. $\bm{115}$, 036807 (2015).}
\bibitem{H. Huang 2016}
H. Huang, J. Liu, D. Vanderbilt, and W. Duan,
\href{https://journals.aps.org/prb/abstract/10.1103/PhysRevB.93.201114}{Phys. Rev. B $\bm{93}$, 201114(R) (2016).}
\bibitem{Y. Du 2017}
Y. Du, F. Tang, D. Wang, L. Sheng, E. Kan, C. Duan, S. Y. Savrasov, and X. Wan,
\href{https://www.nature.com/articles/s41535-016-0005-4}{npj Quantum Materials $\bm{2}$, 3 (2017).}
\bibitem{X. Zhang 2017}
X. Zhang , L. Jin, X. Dai, and G. Liu,
\href{https://pubs.acs.org/doi/10.1021/acs.jpclett.7b02129}{Phys. Chem. Lett. $\bm{8}$, 4814-4819 (2017).}
\bibitem{Z. Wang 2017}
Z. Wang and G. Wang,
\href{https://www.sciencedirect.com/science/article/pii/S0375960117306217?via\%3Dihub}{Phys. Lett. A $\bm{381}$, 2856-2859 (2017).}
\bibitem{Q. Xu 2017}
Q. Xu, R. Yu, Z. Fang, X. Dai, and H. Weng,
\href{https://journals.aps.org/prb/abstract/10.1103/PhysRevB.95.045136}{Phys. Rev. B $\bm{95}$, 045136 (2017).}

\bibitem{S. Nie 2018}
S. Nie, H. Weng, and F. B. Prinz,
\href{https://arxiv.org/abs/1803.08486}{arXiv:1803.08486}

\bibitem{G. Bian 2016}
G. Bian, T. Chang, R. Sankar, S. Xu, H. Zheng, T. Neupert, C. Chiu, S. Huang, G. Chang, I. Belopolski, D. S. Sanchez, M. Neupane, N. Alidoust, C. Liu, B. Wang, C. Lee, H. Jeng, C. Zhang, Z. Yuan, S. Jia, A. Bansil, F. Chou, H. Lin, and M. Z. Hasan,
\href{https://www.nature.com/articles/ncomms10556}{Nat. Commun. $\bm{7}$, 10556 (2016).}


\bibitem{J. Lu 2015}
J. Lu, W. Luo, X. Li, S. Yang, J. Cao, X. Gong, and H. Xiang,
\href{http://iopscience.iop.org/article/10.1088/0256-307X/34/5/057302/meta}{Chinese Phys. Lett. $\bm{34}$, 045108 (2015).}
\bibitem{Y. Jin 2017}
Y. Jin, R. Wang, J. Zhao, Y. Du, C. Zheng, L. Gan, J. Liu, H. Xu, and  S. Y. Tong,
\href{http://pubs.rsc.org/en/Content/ArticleLanding/2017/NR/C7NR03520A#!divAbstract}{Nanoscale $\bm{9}$, 13112-13118 (2017).}
\bibitem{P. Zhou 2018}
P. Zhou, Z. S. Ma, and L. Z. Sun,
\href{http://pubs.rsc.org/en/content/articlehtml/2018/TC/C7TC05095J}{Mater. Chem. C $\bm{6}$, 1206-1213 (2018).}

\bibitem{B. Feng 2017}
B. Feng, B. Fu, S. Kasamatsu, S. Ito, P. Cheng, C. Liu, Y. Feng, S. Wu, S. K. Mahatha, P. Sheverdyaeva, P. Moras, M. Arita, O. Sugino, T. Chiang, K. Shimada, K. Miyamoto, T. Okuda, K. Wu, L. Chen, Y. Yao, and  I. Matsuda,
\href{https://www.nature.com/articles/s41467-017-01108-z}{Nat. Commun. $\bm{8}$, 1007 (2017).}

\bibitem{S. Park 2017}
S. Park and B.-J. Yang,
\href{https://journals.aps.org/prb/abstract/10.1103/PhysRevB.96.125127}{Phys. Rev. B $\bm{96}$, 125127 (2017).}


\bibitem{B. Yang 2016}
B. Yang, X. Zhang, and M. Zhao,
\href{https://arxiv.org/abs/1612.06562}{arXiv:1612.06562}
\bibitem{C. Niu 2017}
C. Niu, P. M. Buh, H. Zhang, G. Bihlmayer, D. Wortmann, S. Bl\"ugel, and Y. Mokrousov,
\href{https://arxiv.org/abs/1703.05540}{arXiv:1703.05540}



\bibitem{G. B. Liu 2013}
G-. B. Liu, W-. Y. Shan, Y. Yao, W. Yao, and and D. Xiao,
\href{https://journals.aps.org/prb/abstract/10.1103/PhysRevB.88.085433}{Phys. Rev. B $\bm{88}$, 085433 (2013).}
\bibitem{H. Rostami 2015}
H. Rostami, R. Rold\'an, E. Cappelluti, R. Asgari, and F. Guinea,
\href{https://journals.aps.org/prb/abstract/10.1103/PhysRevB.92.195402}{Phys. Rev. B $\bm{92}$, 195402 (2015).}


\bibitem{K. Kosmider 2013}
K. Ko\'smider, J. W. Gonz\'alez, and J. Fern\'andez-Rossier,
\href{https://journals.aps.org/prb/abstract/10.1103/PhysRevB.88.245436}{Phys. Rev. B $\bm{88}$, 245436 (2013).}

\bibitem{A. Kormanyos 2014}
A. Korm\'anyos, V. Z\'olyomi, N. D. Drummond, and G. Burkard,
\href{https://journals.aps.org/prx/abstract/10.1103/PhysRevX.4.011034}{Phys. Rev. X $\bm{4}$, 011034 (2014).}

\bibitem{A. Kormanyos 2015}
A. Korm\'anyos, G. Burkard, M. Gmitra, J. Fabian, V. Z\'olyomi, N. D. Drummond, and V. I. Fal'ko,
\href{http://iopscience.iop.org/article/10.1088/2053-1583/2/2/022001/meta}{2D Mater. $\bm{2}$, 022001 (2015).}

\bibitem{J. P. Echeverry 2016}
J. P. Echeverry, B. Urbaszek, T. Amand, X. Marie, and I. C. Gerber,
\href{https://journals.aps.org/prb/abstract/10.1103/PhysRevB.93.121107}{Phys. Rev. B $\bm{93}$, 121107(R) (2016).}



\bibitem{M. Koperski 2017} 
M. Koperski, M. R. Molas, A. Arora, K. Nogajewski, A. O. Slobodeniuk, C. Faugeras, and M. Potemski,
\href{https://www.degruyter.com/view/j/nanoph.2017.6.issue-6/nanoph-2016-0165/nanoph-2016-0165.xml}{Nanophotonics $\bm{6}$, 1289-1308 (2017).}

\bibitem{Z. Ye 2014} 
Z. Ye, T. Cao, K. O’Brien, H. Zhu, X. Yin, Y. Wang, S. G. Louie, and X. Zhang,
\href{https://www.nature.com/articles/nature13734}{Nature $\bm{513}$, 214-218 (2014).}

\bibitem{A. Arora 2015} 
A. Arora, K. Nogajewski, M. Molas, M. Koperski, and M. Potemski,
\href{https://pubs.rsc.org/en/content/articlelanding/2015/nr/c5nr06782k#!divAbstract}{Nanoscale $\bm{7}$, 20769-20775 (2015).}

\bibitem{G. Wang 2015} 
G. Wang, C. Robert, A. Suslu, B. Chen, S. Yang, S. Alamdari, I. C. Gerber, T. Amand, X. Marie, S. Tongay, and B. Urbaszek,
\href{https://www.nature.com/articles/ncomms10110}{Nat. Comm. $\bm{6}$, 10110 (2015).}

\bibitem{M. R. Molas 2017} 
M. R. Molas, C. Faugeras, A. O. Slobodeniuk, K. Nogajewski, M. Bartos, D. M. Basko, and M. Potemski ,
\href{http://iopscience.iop.org/article/10.1088/2053-1583/aa5521}{2D Materials $\bm{4}$, 021003 (2017).}

\bibitem{X. X. Zhang 2017} 
X-. X. Zhang, T. Cao, Z. Lu, Y-. C. Lin, F. Zhang, Y. Wang, Z. Li, J. C. Hone, J. A. Robinson, D. Smirnov, S. G. Louie, and T. F. Heinz,
\href{https://www.nature.com/articles/nnano.2017.105}{Nat. Nanotech. $\bm{12}$, 883-888 (2017).}


\bibitem{A. Kormanyos 2013}
A. Korm\'anyos, V. Z\'olyomi, N. D. Drummond, P. Rakyta, G. Burkard, and V. I. Fal'ko,
\href{https://journals.aps.org/prb/abstract/10.1103/PhysRevB.88.045416}{Phys. Rev. B $\bm{88}$, 045416 (2013).}



\bibitem{K. Dolui 2013}
K. Dolui, I. Rungger, and S. Sanvito,
\href{https://journals.aps.org/prb/abstract/10.1103/PhysRevB.87.165402}{Phys. Rev. B $\bm{87}$, 165402 (2013).}

\bibitem{D. Sercombe 2013}
D. Sercombe, S. Schwarz, O. D. Pozo-Zamudio, F. Liu, B. J. Robinson, E. A. Chekhovich, I. I. Tartakovskii, O. Kolosov, and A. I. Tartakovskii,
\href{https://www.nature.com/articles/srep03489}{Sci. Rep. $\bm{3}$, 3489 (2013).}

\bibitem{K. Kim 2018}
K. H. Kim and H. W. Lee,
\href{https://journals.aps.org/prb/abstract/10.1103/PhysRevB.97.235423}{Phys. Rev. B $\bm{97}$, 235423 (2018).}
\bibitem{B. T. Zhou 2017}
B. T. Zhou, K. Taguchi, Y. Kawaguchi, Y. Tanaka, and K. T. Law, 
\href{https://arxiv.org/abs/1712.02942}{arXiv:1712.02942.}
\bibitem{A. S. Patri 2018}
A. S. Patri, K. Hwang, H. Lee, and Y. B. Kim,
\href{https://www.nature.com/articles/s41598-018-26355-y}{Sci. Rep. $\bm{8}$, 8052 (2018).}

\bibitem{VESTA}
K. Momma and F. Izumi,
\href{http://scripts.iucr.org/cgi-bin/paper?S0021889811038970}{J. Appl. Crystallogr. $\bm{44}$, 1272-1276 (2011).}












\end{thebibliography}
\end{document}